\documentclass[12pt]{article}
\usepackage{amssymb,amscd}
\title{Integrable and non-integrable equations with peakons} 

\author{A. Degasperis\thanks{Dipartimento di Fisica,
Universit\`{a} di Roma ``La Sapienza'', P.le A. Moro 2,
00185 Roma, Italia. E-mail:
antonio.degasperis@roma1.infn.it},
D.D. Holm\thanks{Theoretical Division and Center for
Nonlinear Studies, Los Alamos National Laboratory, Los
Alamos, NM 87545, USA. E-mail:dholm@lanl.gov}
\& A.N.W. Hone\thanks{Institute of Mathematics \& Statistics,
University of Kent, Canterbury CT2 7NF, UK. 
E-mail:anwh@ukc.ac.uk}
}
\begin{document}
\renewcommand{\theequation}{\arabic{section}.\arabic{equation}}
\newcommand{\beq}{\begin{equation}}
\newcommand{\eeq}{\end{equation}}
\newcommand{\bea}{\begin{eqnarray}}
\newcommand{\eea}{\end{eqnarray}}
\maketitle

\begin{abstract}
We consider a one-parameter family of non-evolutionary 
partial differential equations which includes the 
integrable Camassa-Holm equation and a new integrable 
equation first isolated by Degasperis and Procesi. 
A Lagrangian and Hamiltonian formulation is presented 
for the whole family of equations, and we discuss how 
this fits into a bi-Hamiltonian framework in the integrable 
cases. The Hamiltonian dynamics of peakons and some 
other special finite-dimensional reductions are      
also described.  
\end{abstract} 

\section{Introduction} 
In this note we consider the following one-parameter 
family of partial differential equations (PDEs): 
\beq 
u_t-u_{xxt}+(b+1)uu_x=bu_xu_{xx}+uu_{xxx}  \label{eq:bfamily}
\eeq
(the parameter $b$ is constant). 
In the particular case $b=2$ this becomes the dispersionless 
version of the integrable 
Camassa-Holm equation, which is  
\beq
u_t-u_{xxt}+3uu_x=2u_xu_{xx}+uu_{xxx}. 
\label{eq:caholm}
\eeq
With the inclusion of linear dispersion terms the equation 
(\ref{eq:caholm}) takes the form   
\beq 
u_t +c_0 u_x + \gamma u_{xxx} -\alpha^2 u_{xxt} 
=2u_xu_{xx}+uu_{xxx}  
\label{eq:disp} 
\eeq 
($c_0,\gamma$ and $\alpha$ are constant parameters), 
and in this form it was derived  
as an approximation to the incompressible Euler
equations, and found to be completely integrable with a
Lax pair and associated bi-Hamiltonian structure \cite{ch,ch2}. 

Degasperis and Procesi \cite{dega} 
applied the method of 
asymptotic integrability to test a many-parameter family 
of equations generalizing  
(\ref{eq:disp}), and found that only three equations  
passed the test up to third order in the asymptotic expansion, 
namely KdV, Camassa-Holm and one new equation. After 
rescaling and applying a Galilean transformation, the new 
equation may be written in dispersionless form as 
\beq
u_t-u_{xxt}+4uu_x=3u_xu_{xx}+uu_{xxx}. \label{eq:tdnodisp}
\eeq
Henceforth we refer to (\ref{eq:tdnodisp}) as the 
Degasperis-Procesi equation, and observe that it 
is the $b=3$ case of the equation (\ref{eq:bfamily}). 
In a recent article \cite{needs} we proved the integrability 
of the Degasperis-Procesi equation by providing a Lax 
pair, and derived two infinite sequences of conservation laws. 
The key to our construction 
was to use a reciprocal transformation connecting 
the equation (\ref{eq:tdnodisp}) with a negative 
flow in the Kaup-Kupershmidt hierarchy. 

Both the Camassa-Holm equation (\ref{eq:caholm}) and the 
Degasperis-Procesi equation (\ref{eq:tdnodisp}) display 
the weak Painlev\'{e} property \cite{weak}, with 
algebraic branching in their solutions. Thus when Gilson 
and Pickering applied the standard 
Painlev\'{e} tests to a class of equations including 
the family (\ref{eq:bfamily}), these two integrable 
equations were excluded because of the branching, 
and they found that no equations 
passed the tests. In a forthcoming work \cite{longer} 
we show that after applying a reciprocal transformation 
to the family (\ref{eq:bfamily}) it is possible to use 
Painlev\'{e} analysis effectively, and the requirement of 
only pole singularities immediately isolates the 
two integrable cases $b=2,3$. Because the equations 
(\ref{eq:bfamily}) are non-evolutionary, the standard 
symmetry approach of Shabat et al \cite{shabat} does not 
apply. However, recently Mikhailov and Novikov developed 
a powerful extension of the symmetry classification 
method \cite{mik}, and applying this to the equations 
(\ref{eq:bfamily}) they found that only the cases $b=2,3$ 
could possess infinitely many commuting symmetries,  
and so only these two cases are integrable.            

One of the novel properties of the Camassa-Holm 
equation is that it admits exact solutions in terms of 
a superposition of an arbitrary number of 
peakons, or peaked solitons. The $N$-peakon solution is just 
\beq
u(x,t)=\sum_{j=1}^Np_j(t)e^{-|x-q_j(t)|},
\label{eq:Npeakon}
\eeq
where the positions $q_j(t)$ of the peaks and their 
momenta $p_j(t)$ satisfy an associated dynamical 
system taking the canonical Hamiltonian form 
\beq
\dot{p}_j =
-\,\frac{\partial h}{\partial q_j},
\qquad
\dot{q}_j  =
\frac{\partial h}{\partial p_j}
,
\label{eq:chODEeqn}
\eeq
with 
\beq 
h=\frac{1}{2}\sum_{j,k=1}^{N} p_j p_k\, e^{-|q_j-q_k|}. 
\label{eq:cham} 
\eeq 
It is shown in \cite{ch, ch2} that this is a 
completely integrable finite-dimensional system, with 
an $N\times N$ Lax pair. The integrability of 
the Camassa-Holm peakon dynamics further received an 
r-matrix interpretation in \cite{rabru}. 
 
One of our observations in \cite{needs} was that all 
of the equations in the family (\ref{eq:bfamily}) 
admit $N$-peakon solutions in the form (\ref{eq:Npeakon}), 
but the dynamical system for $q_j$, $p_j$ 
only takes the canonical Hamiltonian form 
(\ref{eq:chODEeqn})  
in the particular case $b=2$. In the following we 
construct a Lagrangian (for $b\neq 0$) 
and associated Hamiltonian 
structure, depending on the parameter $b$, 
for each equation in the family (\ref{eq:bfamily}). 
We then reduce this onto the finite-dimensional 
submanifold of peakon solutions for each $N$ to obtain 
a new non-canonical Poisson structure and associated Hamiltonian 
form for the peakon dynamics. It turns out the two-body 
dynamics ($N=2$) is Liouville 
integrable for any value of $b$, but 
for arbitrary $N$ we expect integrability only for 
$b=2,3$ since these are the reductions of 
integrable PDEs.  An $N\times N$ matrix Lax matrix 
pair for the $b=3$ peakon dynamics was presented in 
\cite{needs}, but at that stage we were still 
ignorant of the correct Poisson structure.

\section{Lagrangian form and Hamiltonian operator} 

In the following we consider solutions of 
(\ref{eq:bfamily}) on the whole $x$-axis vanishing at 
infinity, and all integrals will be assumed to be 
between $\pm\infty$, although (with suitable modification) 
most of the results hold for solutions with constant background 
or on a periodic domain. Some of the operations below 
only make sense for sufficiently smooth real functions 
$u$, but later we will want to apply some formulae 
to weak solutions such as the peakons, in which case 
we will find that certain manipulations are valid 
also for distributions. 

It is convenient to introduce the quantity 
$$ 
m=u-u_{xx}, 
$$ 
which is just the Helmholtz operator acting on $u$, 
so that the family of equations (\ref{eq:bfamily}) 
may be rewritten in the form 
\beq 
m_t+um_x+b\,u_xm=0. \label{eq:mb} 
\eeq 
For any value of $b$ this is a conservation law, i.e. 
$$
m_t+\left(\frac{(b-1)}{2}(u^2-u_x^2)+um\right)_x=0, 
$$ 
while for all $b\neq 0$ there are at least three 
conserved quantities, namely 
\beq 
\int m\, dx, \quad \int m^{1/b}\, dx, \quad 
\int m^{-1/b}\left(\frac{m_x^2}{b^2m^2}+1\right)\, dx. 
\label{eq:conserv} 
\eeq 
For the integrable cases $b=2$ \cite{ch,ch2,schiff} 
and $b=3$ \cite{needs}  
there are infinitely many conserved quantities.     
 
For the second quantity appearing in (\ref{eq:conserv}) 
above, 
the associated conservation law is conveniently written as 
\beq 
p_t+(pu)_x=0, \qquad m=-p^b. 
\eeq 
With this further rewriting of the equation 
(\ref{eq:mb}), or equivalently (\ref{eq:bfamily}),  
for $b\neq 0$ we 
introduce a potential $\eta$ such that 
$$ 
\eta_x=p, \qquad \eta_t=-pu. 
$$ 
Then in terms of this potential we find that (apart 
from the particular cases $b=0,1$)   
the equation (\ref{eq:bfamily}) may be derived 
from the following action: 
\beq 
S\equiv\int\int\mathcal{L}\, dx\, dt = 
\int\int\left( 
\frac{1}{2}\frac{\eta_t}{\eta_x}\left[ 
(\log \eta_x)_{xx}+1\right] 
+\frac{\eta_x^b}{b-1}\right)dx\, dt. 
\label{eq:lagr} 
\eeq 
When $b=1$ a similar Lagrangian formulation is valid, 
but the Lagrange density must be modified to 
$$ 
\mathcal{L}=\frac{1}{2}\eta_t\eta_x^{-1}\left[ (\log \eta_x)_{xx}+1\right] +\eta_x\log \eta_x. 
$$      
    
Starting from the Lagrangian we can apply a Legendre 
transformation in the usual way. The conjugate momentum 
is 
$$ 
\zeta\equiv\frac{\partial \mathcal{L}}{\partial\eta_t}= 
\frac{1}{2\eta_x}\left[ (\log \eta_x)_{xx}+1
\right], 
$$ 
and (for $b\neq 1$) the Hamiltonian is 
\beq 
H=\int(\zeta\eta_t-\mathcal{L})\, dx=-\frac{1}{b-1}\int 
\eta_x^b\, dx=\frac{1}{b-1}\int m\, dx. 
\label{eq:canham} 
\eeq 
Having applied the Legendre transformation we then find that 
for any $b$ the PDE (\ref{eq:mb}) can be 
written in Hamiltonian form as 
\beq 
m_t=\hat{B}\frac{\delta H}{\delta m}, 
\label{eq:hampde} 
\eeq 
where the operator 
\beq 
\hat{B}=-(bm\partial_x+m_x)(\partial_x-\partial_x^3)^{-1} 
(bm\partial_x+(b-1)m_x) 
\label{eq:hamop} 
\eeq 
is skew-symmetric and satisfies the Jacobi identity 
(for a proof see \cite{longer}). In the case $b=1$ 
the Legendre transformation gives a different result: 
$\int m\, dx$   is a Casimir for the operator $\hat{B}$, 
and instead the Hamiltonian may be taken as 
$$ 
H=\int m\log m\, dx. 
$$ 
The formula (\ref{eq:hampde}) is even valid 
in the case $b=0$, when the Lagrangian formulation breaks 
down. 
  
For the integrable case $b=2$, the 
Camassa-Holm equation (\ref{eq:caholm}), there are two local 
Hamiltonian structures \cite{ch,ch2}, given by 
$$ 
B_0=-\partial_x(1-\partial_x^2), \qquad 
B_1=-(m\partial_x+\partial_x m), 
$$ 
and the compatible operator (\ref{eq:hamop}) 
is the first nonlocal Hamiltonian structure obtained 
by applying the recursion operator $R=B_1 B_0^{-1}$ to 
$B_1$, i.e. 
$$ 
B_2\equiv B_1 B_0^{-1}B_1=\hat{B}|_{b=2}. 
$$ 
In the other integrable case $b=3$, namely 
the Degasperis-Procesi equation (\ref{eq:tdnodisp}), 
there is only one local Hamiltonian structure, and  
the operator (\ref{eq:hamop}) gives the second Hamiltonian        
structure, viz 
\beq 
B_0=-\partial_x(1-\partial_x^2)(4-\partial_x^2), 
\qquad B_1=\hat{B}|_{b=3}. 
\label{eq:tdops} 
\eeq 
This compatible bi-Hamiltonian pair first 
appeared in \cite{needs}. 
The nonlocal part   
of (\ref{eq:hamop}) may be defined precisely
via 
\beq 
(\partial_x-\partial_x^3)^{-1}f(x)=\frac{1}{2} 
\int_{-\infty}^{\infty}G(x-y)f(y)\, dy, \qquad 
G(x)=sgn(x)(1-e^{-|x|}). 
\label{eq:kerdef} 
\eeq 
The Jacobi identity for (\ref{eq:hamop}) reduces 
to a single functional equation satisfied by the 
kernel $G$ \cite{longer}. 

\section{Peakon dynamics and special solutions} 

The peakon solutions of (\ref{eq:bfamily}) take the form 
(\ref{eq:Npeakon}) for any $b$, in which case the 
dependent variable $m$ is 
a sum of Dirac delta functions,   
\beq 
m(x,t)=2\sum_{j=1}^Np_j(t)\delta(x-q_j(t)). 
\label{eq:deltas} 
\eeq 
Introducing the even kernel (Green's function of the 
Helmholtz operator) 
$$ 
g(x)=e^{-|x|}, \qquad G(x)=\int_0^xg(s)\, ds, 
$$ 
the dynamical system for the peakons may be conveniently 
written as 
\beq
\begin{array}{ccl}
\frac{dq_j}{dt} & = &  
\sum_{k=1}^N p_kg(q_j-q_k), \\ 
&& \\   
\frac{dp_j}{dt} & = &  
-(b-1)\sum_{k=1}^N p_jp_kg'(q_j-q_k).
\end{array}
\label{eq:peakdyn}
\eeq
Clearly this takes the canonical Hamiltonian form 
(\ref{eq:chODEeqn}) only for $b=2$. 

In order to obtain the correct Poisson structure 
on the reduced peakon phase space, for any $b\neq 1$ 
we can use the Poisson bracket defined by 
(\ref{eq:hamop}), which gives (up to rescaling) 
$$
 \{ m(x),m(y) \} =\frac{1}{(b-1)}\Big(
G(x-y)m_x(x)m_x(y)+bG'(x-y) m(x)m_x(y) 
$$
\beq
-bG'(x-y)m(y)m_x(x) - b^2 G''(x-y)m(x)m(y)\Big) .
\label{eq:pb}
\eeq 
Substituting the peakon expression (\ref{eq:deltas}) 
into (\ref{eq:pb}), we are able to calculate  
the Poisson bracket on the reduced ($2N$-dimensional) 
phase space as 
$$
\{ p_j,p_k \}=-(b-1)G''(q_j-q_k)p_jp_k, \quad
\{ q_j,p_k \}=p_k G'(q_j-q_k),
$$
\beq 
\{ q_j,q_k \}=1/(b-1) G(q_j-q_k).
\label{eq:fibracket}
\eeq 
The Jacobi identity for this 
non-canonical Poisson bracket  
follows from the same functional equation for $G$ 
that arises for the operator (\ref{eq:hamop}) 
(see \cite{longer}).  The Hamiltonian for the dynamical system 
(\ref{eq:peakdyn}) is found simply by substituting 
(\ref{eq:deltas}) into the PDE Hamiltonian 
(\ref{eq:canham}), so that after scaling we have 
$$ 
\tilde{h}=(b-1)H/2=\sum_{j=1}^Np_j. 
$$ 

Having obtained the reduction of the  Poisson structure 
for the PDE, we see that the peakon dynamical 
system (\ref{eq:peakdyn}) takes the Hamiltonian form    
$$
\frac{dq_j}{dt}=\{ q_j,\tilde{h}\}, 
\qquad \frac{dp_j}{dt}=\{ p_j,\tilde{h}\}.  
$$
Now for any $b$ we also introduce the second quantity 
$$ 
k=\frac{1}{2}\sum_{j,k=1}^N 
p_jp_k\Big(1-(1-g(q_j-q_k))^{b-1}\Big). 
$$ 
For $b=2$ this just coincides with the canonical 
Hamiltonian (\ref{eq:cham}) for $N$ Camassa-Holm peakons, 
$k=h$, and is clearly conserved. For the Degasperis-Procesi 
case $b=3$, with the bracket (\ref{eq:fibracket}) 
$k$ Poisson commutes with $\tilde{h}$ for any $N$; 
it is the second conserved quantity obtained from 
${\tt trace}\,\mathsf{L}^2$, where $\mathsf{L}$ is the Lax matrix 
presented in \cite{needs}. However, for any other 
value of $b$ it seems that $k$ is conserved only in the 
two-body case $N=2$. Thus we see that for generic 
values of the parameter $b$  
only the two peakon dynamics is Liouville integrable. 
The two-body dynamics and associated peakon phase 
shifts are described in more detail in \cite{longer}.      
  
The integrability of the two-body dynamics seems to 
induce remarkable stability properties for the peakons 
even in the non-integrable regime. Numerical studies 
of the PDE family 
(\ref{eq:bfamily}) on a periodic domain \cite{hs} show that 
starting from a Gaussian initial profile 
for $u$, for any $b>1$ a train of peakons emerge and preserve  
their shape and speed after interaction, undergoing only a 
phase shift. The peakons are unstable for $b<1$. 
For the range $-1<b<1$ the numerical 
results indicate that ``ramps-and-cliffs'' profiles, like those 
appearing as solutions to the Burgers equation, 
seem to be stable. We do not have an exact analytic 
expression for such profiles. However, the ``ramp'' 
part behaves like the exact similarity solution 
$$ 
u(x,t)=\frac{x}{(b+1)t}, 
$$ 
while for the special case $b=0$ we find an exact 
solution in terms of a superposition of $N$ ``cliffs,'' 
$$ 
u(x,t)=c+\sum_{j=1}^N a_j G(x-ct-Q_j(t)), 
$$     
with $c$ and the $a_j$ being constants. The 
$Q_j$ satisfy the Hamiltonian dynamical system 
$$ 
\frac{dQ_j}{dt}=\sum_{k\neq j}a_k G(Q_j-Q_k)\equiv \{Q_j,\hat{h}\},
\qquad \hat{h}=\sum_{j=1}^Na_jQ_j,
$$ 
where (after scaling) the Poisson bracket reduced from
(\ref{eq:pb}) is
$$
\{ Q_j,Q_k\}=G(Q_j-Q_k).
$$
The ``cliff'' dynamics is integrable for $N=2$ and $N=3$.

In the remaining range $b<-1$, the numerical results of 
\cite{hs} show that  Gaussian initial data               
splits into a superposition of pulses which are 
asymptotically stationary. For that parameter range 
there is a unique vanishing stationary solution 
given by the exact sech-shaped profile 
$$ 
u(x)=\Big(\cosh [(b+1)(x-x_0)/2] \Big)^{2/(b+1)}  
$$    
centred around the arbitrary point $x_0$. 
In future work we intend to address the stability properties of
the peakons and other exact solutions.

\noindent 
{\bf Acknowledgements:} AH would like to thank the organizers of Nonlinear 
Physics II and the IMS (University of Kent) for supporting his attendance 
at the Gallipoli meeting.

\end{document}